\documentclass{article}
\usepackage{amsmath,amssymb}
\begin{document}

\begin{center}
{\Large\bf Fundamental solutions for a class of three-dimensional elliptic
equations with singular coefficients}

\bigskip

\bf{Anvar Hasanov and E.T. Karimov}


\bigskip

\emph{Institute of Mathematics and Information Technologies, Uzbek
Academy of Sciences, Durmon yuli str. 29, Tashkent 100125,
Uzbekistan,}\\ E-mail: anvarhasanov@yahoo.com, \,\,\,erkinjon@gmail.com
\end{center}

\bigskip

\begin{abstract}
We consider an equation
$$
L_{\alpha ,\beta ,\gamma } \left( u \right) \equiv u_{xx}  +
u_{yy}  + u_{zz}  + \displaystyle \frac{{2\alpha }}{x}u_x +
\displaystyle \frac{{2\beta }}{y}u_y  + \displaystyle
\frac{{2\gamma }}{z}u_z  = 0
$$
in a domain ${\bf R}_3^ +   \equiv \left\{ {
\left ( {x,y,z} \right):\,x
> 0,\,y > 0,\,z > 0} \right\}$. Here $\alpha ,\beta ,\gamma$ are constants, moreover \\
$0 < 2\alpha ,2\beta
,2\gamma  < 1$. Main result of this paper is a construction of eight fundamental solutions for above-given equation in an explicit form. They are expressed by Lauricella's hypergeometric functions of three variables. Using expansion of Lauricella's hypergeometric function by products of
Gauss's hypergeometric functions, it is proved that the found solutions have a singularity of the order $1/r$ at $r \to 0$. Furthermore, some properties of these solutions, which will be used at solving boundary-value problems for afore-mentioned equation are shown.
\end{abstract}

\textbf{Keywords:} Fundamental solutions, Elliptic differential equation with singular coefficients, Lauricella's hypergeometric
function of three variables.

\textbf{MSC 2000:} \emph{Primary} 35A08 \emph{Secondary} 35J70

\section{Introduction}

Known that fundamental solutions (FSs) have an essential role at studying partial differential equations (PDEs). Formulation and solving of many local and non-local boundary-value problems (BVPs) are based on these solutions. Moreover, FSs appear as potentials, for instance, as a simple-layer and double-layer potentials in the theory of potentials.

Explicit form of FSs gives a possibility to study considered equation in detail. For example, in the works [3-5] by J.Barros-Neto and I.M.Gelfand, FSs for Tricomi operator, relative to an arbitrary point in the plane were explicitly calculated. Shown that found FSs clearly reflect the change of type of the Tricomi operator across the $x$-axis. We also mention J.Leray's work [12], where a general method,
based upon the theory of analytic functions of several complex variables, for finding FSs for a class of hyperbolic linear differential operators with analytic coefficients was described. In particular, he showed how his method could be used to obtain, in the hyperbolic region, a FS for the Tricomi operator relative to a point $(0, b)$. Among other results on this direction we would like to note work by M.Itagaki [11], where three-dimensional high-order fundamental solutions for modified Helmholtz equation were found. Found solutions can be applied with the boundary particle method to some 2D inhomogeneous problems, for example, see [8].

Various modifications of the equation
\begin{equation}
L_{\alpha ,\beta ,\gamma } \left( u \right) \equiv u_{xx} +
u_{yy}  + u_{zz}  + \displaystyle \frac{{2\alpha }}{x}u_x +
\displaystyle \frac{{2\beta }}{y}u_y  + \displaystyle
\frac{{2\gamma }}{z}u_z = 0,\,\,0 < 2\alpha ,2\beta ,2\gamma  <
1,\,\,\alpha ,\beta ,\gamma = const,
\end{equation}
in two-dimensional case were considered in many papers [1, 7, 13-18].
For example, in [7] the author is concerned with
generalized bi-axially symmetric potentials (GBSP), i.e., with
functions

$$
u\left( {x,y} \right) = \sum\limits_{n = 0}^\infty \displaystyle
{} a_n r^{2n} P_n^{\left( {\alpha ,\beta } \right)} \left( {\cos
2\theta } \right),
$$
where $x = r\cos \theta ,\,y = r\sin \theta ,\,\alpha ,\beta  > -
1/2,\,P_n^{\left( {\alpha ,\beta } \right)} \left( {\cos 2\theta }
\right)$ are Jacobi polynomials. These functions occur as
solutions of the equation
$$
u_{xx}  + u_{yy}  + \displaystyle \frac{{2\beta  + 1}}{x}u_x +
\displaystyle \frac{{2\alpha + 1}}{y}u_y  = 0.
$$
Theorems on the convergence of above series and
growth properties are obtained in terms of the coefficients $a_n$.

In [1], the elliptic or ultra hyperbolic equation
\begin{equation}
Lu = \displaystyle \sum\limits_{i = 1}^n {} \left(
{\frac{{\partial ^2 u}}{{\partial x_i^2 }} + \displaystyle
\frac{{\alpha _i }}{{x_i }}\displaystyle \frac{{\partial
u}}{{\partial x_i }}} \right) \pm \displaystyle \sum\limits_{i =
1}^s {} \left( {\displaystyle \frac{{\partial ^2 u}}{{\partial
y_i^2 }} + \displaystyle \frac{{\beta _i }}{{x_i }}\displaystyle
\frac{{\partial u}}{{\partial y_i }}} \right) + \displaystyle
\frac{{\gamma u}}{{r^2 }} = 0,
\end{equation}
is considered, where the constants $\alpha _1 ,...,\alpha _n
,\,\beta _1 ,...,\beta _s \,$ $\gamma$ are real parameters, $r^2  = \sum\limits_{i = 1}^n {} x_i^2  \pm \sum\limits_{i = 1}^s
{} y_i^2$ and establishes some properties of the operator $L$. An
example is the Kelvin principle for the equation (2): If
$u\left( {x,y} \right)$ is a solution of (2), then the function
$\sigma  = r^{ - \varphi } u\left( {\xi ,\eta } \right)$ is also a
solution of the same equation, where $\xi _i  = x_i /r^2 ,\,\eta
_i  = y_i /r^2 ,\,\varphi  = n + s - 2 + \sum\limits_{i = 1}^n {}
\alpha _i  + \sum\limits_{i = 1}^s {} \beta _i$. For a class of
iterated equations $L^p u = 0$, it was shown that if $Lu = 0$,
then $L^p \left[ {r^{2j} u\left( {x,y} \right)} \right] = L^p
\left[ {r^{2j - \varphi } u\left( {\xi ,\eta } \right)} \right] =
0$ for $j = 0,...,p - 1;$ if, the function $u\left( {x,y} \right)$
 is a homogeneous solution of degree $\lambda$ of the equation $Lu =
 0$, then $L^p \left[ {r^{2j - \varphi  - 2\lambda } u\left( {x,y}
\right)} \right] = L^p \left[ {r^{2\left( {j + \lambda } \right)}
u\left( {\xi ,\eta } \right)} \right] = 0$ for     $j = 0,1,...,p
- 1.$ \\

FSs for elliptic equations with singular coefficients are directly connected with hypergeometric functions (HFs). Therefore, basic properties such as decomposition formulas, integral representations, formulas of analytical continuation, formulas of differentiation for HFs are necessary for studying FSs.

Since afore-mentioned properties of HFs of Gauss, Appell, Kummer were well-known [2], results on investigations of elliptic equations with one or two singular coefficients were successful.
Up to publication results on Lauricella HFs by A.Hasanov and H.M.Srivastava [9, 10] there was no possibility to investigate FSs for the equation (1), since their consist Lauricella HFs of three variables.

In the present paper we construct eight fundamental solutions for the equation (1) in an explicit form. Using decomposition formulas for Lauricella HFs of three variables by simple Gauss HFs, we prove that found FSs have a singularity of order $1/r$ at $r\rightarrow 0$.

The plan of this paper is as follows. In Section 2 we briefly give some preliminary information, which will be used further. Also some constructive formulas for the operator $L_{\alpha,\beta,\gamma}$ are given. In section 3 we describe the method of finding FSs for considered equation and show what order of singularity will have found solutions.

\section{Preliminaries}
Below we give some formulas for Euler gamma-function, Gauss HF, Lauricella HF of three variables, which will be used in next section.

Known that Euler gamma-function $\Gamma(a)$ has properties ([6], pp.17-19, (2), (10), (15))
\begin{equation}
\Gamma(a+m)=\Gamma(a) (a)_m;\,\,\, \Gamma(a+1/2)=\frac{\sqrt{\pi}\Gamma(2a)}{2^{2a-1}\Gamma(a)},\,\, \Gamma(1/2)=\sqrt{\pi}.
\end{equation}
Here $(a)_m$ is a Pochgammer symbol, for which an equality $(a)_{m+n}=(a)_m (a+m)_n$ is true ([6], p.67, (5)).

A function $_2F_1(a,b;c;x)=\displaystyle{\sum\limits_{n=0}^{\infty}\frac{(a)_n(b)_n}{(c)_n n!}x^n}$ is known as Gauss HF and an equality
\begin{equation}
_2F_1(a,b;c;1)=\frac{\Gamma(c)\Gamma(c-a-b)}{\Gamma(c-b)\Gamma(c-a)},\,c\neq 0,-1.,-2,...,\,\,Re(c-a-b)>0
\end{equation}
is hold ([6[, p.73, (14)).
Moreover, the following autotransformer formula ([6], p.76, (22))
\begin{equation}
_2F_1(a,b;c;x)=(1-x)^{-b}\, _2F_1(c-a,b;c;\frac{x}{x-1})
\end{equation}
is valid.

The following system called as the system of hypergeometric equations of Lauricella ([2], p. 117):
\begin{equation}
\left\{ {\begin{array}{*{20}c}
\begin{array}{l}
\xi \left( {1 - \xi } \right)\omega _{\xi \xi }  - \xi \eta \omega
_{\xi \eta }  - \xi \zeta \omega _{\xi \zeta }  + \left[ {2\alpha
- \left( {\alpha  + \beta  + \gamma  + \frac{1}{2} + \alpha  + 1}
\right)\xi } \right]\omega _\xi   - \alpha
\eta \omega _\eta   \\
\,\,\,\,\,\,\,\,\,\,\,\,\,\,\,\,\,\,\,\,\,\,\,\,\,\,\,\,\,\,\,\,\,\,\,\,\,\,\,\,
\,\,\,\,\,\,\,\,\,\,\,\,\,\,\,\,\,\,\,\,\,\,\,\,\,\,\,\,\,\,\,\,\,\,\,\,\,\,\,\,\,\,\,\,
- \alpha \zeta \omega _\zeta   - \left( {\alpha  + \beta  + \gamma
+ \frac{1}{2}} \right)
\alpha \omega  = 0, \\
\end{array}  \\
\begin{array}{l}
\eta \left( {1 - \eta } \right)\omega _{\eta \eta }  - \xi \eta
\omega _{\xi \eta }  - \eta \zeta \omega _{\eta \zeta }  + \left[
{2\beta  - \left( {\alpha  + \beta  + \gamma  +
\frac{1}{2} + \beta  + 1} \right)\eta } \right]\omega _\eta   - \beta \xi \omega _\xi   \\
\,\,\,\,\,\,\,\,\,\,\,\,\,\,\,\,\,\,\,\,\,\,\,\,\,\,\,\,\,\,\,\,\,\,\,\,\,\,\,\,\,\,\,\,\,
\,\,\,\,\,\,\,\,\,\,\,\,\,\,\,\,\,\,\,\,\,\,\,\,\,\,\,\,\,\,\,\,\,\,\,\,\,\,\,
- \beta \zeta \omega _\zeta   - \left( {\alpha  + \beta  + \gamma
+ \frac{1}{2}} \right)\beta
\omega  = 0, \\
\end{array}  \\
\begin{array}{l}
\zeta \left( {1 - \zeta } \right)\omega _{\zeta \zeta }  - \xi
\zeta \omega _{\xi \zeta }  - \eta \zeta \omega _{\eta \zeta }  +
\left[ {2\gamma  - \left( {\alpha  + \beta  + \gamma  +
\frac{1}{2} + \gamma  + 1} \right)\zeta } \right]\omega _\zeta   - \gamma \xi \omega _\xi   \\
\,\,\,\,\,\,\,\,\,\,\,\,\,\,\,\,\,\,\,\,\,\,\,\,\,\,\,\,\,\,\,\,\,\,\,\,\,\,\,\,\,\,\,\,\,\,\,
\,\,\,\,\,\,\,\,\,\,\,\,\,\,\,\,\,\,\,\,\,\,\,\,\,\,\,\,\,\,\,\,\,\,\,\,\,\,
- \gamma \eta \omega _\eta   - \left( {\alpha  + \beta  + \gamma +
\frac{1}{2}} \right)\gamma
\omega  = 0. \\
\end{array}  \\
\end{array}} \right.
\end{equation}
The system (6) has the following
solutions ([2], p. 118):
\begin{equation}
\omega _1 \left( {\xi ,\eta ,\zeta } \right) = F_A^{\left( 3
\right)} \left( {\alpha  + \beta  + \gamma  + \frac{1}{2};\alpha
,\beta ,\gamma ;2\alpha ,2\beta ,2\gamma ;\xi ,\eta ,\zeta }
\right),
\end{equation}
\begin{equation}
\omega _2 \left( {\xi ,\eta ,\zeta } \right) = \xi ^{1 - 2\alpha }
F_A^{\left( 3 \right)} \left( { - \alpha  + \beta  + \gamma  +
\frac{3}{2};1 - \alpha ,\beta ,\gamma ;2 - 2\alpha ,2\beta
,2\gamma ;\xi ,\eta ,\zeta } \right),
\end{equation}
\begin{equation}
\omega _3 \left( {\xi ,\eta ,\zeta } \right) = \eta ^{1 - 2\beta }
F_A^{\left( 3 \right)} \left( {\alpha  - \beta  + \gamma  +
\frac{3}{2};\alpha ,1 - \beta ,\gamma ;2\alpha ,2 - 2\beta
,2\gamma ;\xi ,\eta ,\zeta } \right),
\end{equation}
\begin{equation}
\omega _4 \left( {\xi ,\eta ,\zeta } \right) = \zeta ^{1 - 2\gamma
} F_A^{\left( 3 \right)} \left( {\alpha  + \beta  - \gamma  +
\frac{3}{2};\alpha ,\beta ,1 - \gamma ;2\alpha ,2\beta ,2 -
2\gamma ;\xi ,\eta ,\zeta } \right),
\end{equation}
\begin{equation}
\omega _5 \left( {\xi ,\eta ,\zeta } \right) = \xi ^{1 - 2\alpha }
\eta ^{1 - 2\beta } F_A^{\left( 3 \right)} \left( { - \alpha  -
\beta  + \gamma  + \frac{5}{2};1 - \alpha ,1 - \beta ,\gamma ;2 -
2\alpha ,2 - 2\beta ,2\gamma ;\xi ,\eta ,\zeta } \right),
\end{equation}
\begin{equation}
\omega _6 \left( {\xi ,\eta ,\zeta } \right) = \xi ^{1 - 2\alpha }
\zeta ^{1 - 2\gamma } F_A^{\left( 3 \right)} \left( { - \alpha  +
\beta  - \gamma  + \frac{5}{2};1 - \alpha ,\beta ,1 - \gamma ;2 -
2\alpha ,2\beta ,2 - 2\gamma ;\xi ,\eta ,\zeta } \right),
\end{equation}
\begin{equation}
\omega _7 \left( {\xi ,\eta ,\zeta } \right) = \eta ^{1 - 2\beta }
\zeta ^{1 - 2\gamma } F_A^{\left( 3 \right)} \left( {\alpha  -
\beta  - \gamma  + \frac{5}{2};\alpha ,1 - \beta ,1 - \gamma
;2\alpha ,2 - 2\beta ,2 - 2\gamma ;\xi ,\eta ,\zeta } \right),
\end{equation}
\begin{equation}
\begin{array}{l}
\omega _8 \left( {\xi ,\eta ,\zeta } \right)\\ = \xi ^{1 - 2\alpha
} \eta ^{1 - 2\beta } \zeta ^{1 - 2\gamma }F_A^{\left( 3 \right)}
\left( { - \alpha  - \beta  - \gamma  + \frac{7}{2};1 - \alpha ,1
- \beta ,1 - \gamma ;2 - 2\alpha ,2 - 2\beta ,2 - 2\gamma ;\xi
,\eta ,\zeta } \right),
\end{array}
\end{equation}
where ([2], pp. 114 - 115 (1), (5))
$$
F_A^{\left( 3 \right)} \left( {a;b_1,b_2,b_3;c_1,c_2,c_3;x,y,z}
\right)=\displaystyle \sum \limits_{i,j,k = 0}^\infty
{}\displaystyle\frac{(a)_{i+j+k}(b_1)_i(b_2)_j(b_3)_k}{(c_1)_i(c_2)_j(c_3)_ki!j!k!}x^iy^jz^k,\,\,(|x|+|y|+|z|<1),
$$
$$
\begin{array}{l}
F_A^{\left( 3 \right)} \left( {a;b_1,b_2,b_3;c_1,c_2,c_3;x,y,z}
\right)= \displaystyle \frac{ \Gamma (c_1)\Gamma (c_2)\Gamma
(c_3)}{\Gamma (b_1)\Gamma (b_2)\Gamma (b_3)\Gamma (c_1-b_1)\Gamma
(c_2-b_2)\Gamma (c_3-b_3)} \\
\times \displaystyle \int \limits _0^1\int \limits _0^1\int
\limits _0^1 t_1^{b_1-1}t_2^{b_2-1}t_3^{b_3-1}(1-t_1)^{c_1-b_1-1}
(1-t_2)^{c_2-b_2-1}(1-t_3)^{c_3-b_3-1}(1-xt_1-yt_2-zt_3)^{-a}dt_1dt_2dt_3,\\
Re \, c_i > Re \, b_i >0,\,\,\,i=1,2,3.
\end{array}
$$

The following decomposition formula ([9], p. 117,(14))
\begin{equation}
\begin{array}{l}
\displaystyle F_A^{\left( 3 \right)} \left( {a;b_1 ,b_2 ,b_3 ;c_1
,c_2 ,c_3 ;x,y,z} \right) = \displaystyle \sum\limits_{l,m,n =
0}^\infty  {} \displaystyle{\frac{{\left( a \right)_{l + m + n}
\left( {b_1 } \right)_{l + m} \left( {b_2 } \right)_{l + n} \left(
{b_3 } \right)_{m + n} }}{{ \left( {c_1 } \right)_{l + m} \left(
{c_2 } \right)_{l + n} \left( {c_3 } \right)_{m + n} l!m!n!}}x^{l + m} y^{l + n} z^{m + n}} \\
\times  {}\displaystyle _2F_1 \left( {a + l + m,b_1  + l + m;c_1  +
l + m;x} \right){}\displaystyle _2F_1 \left( {a + l + m + n,b_2  + l + n;c_2  + l + n;y} \right) \\
\times  {}\displaystyle _2F_1 \left( {a + l + m + n,b_3  + m + n;c_3  + m + n;z} \right) \\
\end{array}
\end{equation}
and formula of differentiation ([2], p.19, (20))
\begin{equation}
\frac{{\partial ^{i + j + k} F_A^{(3)} }}{{\partial x^i y^j z^k }} = \frac{{\left( a \right)_{i + j + k} \left( {b_1 } \right)_i \left( {b_2 } \right)_j \left( {b_3 } \right)_k }}{{\left( {c_1 } \right)_i \left( {c_2 } \right)_j \left( {c_3 } \right)_k }}F_A^{(3)} \left( {a + i + j + k,b_1  + i,b_2  + j,b_3  + k;c_1  + i,c_2  + j,c_3  + k;x,y,z} \right)
\end{equation}
are valid.

One of the reasons why we consider the equation (1) is an existing of constructive formulas for the operator $L_{\alpha,\beta,\gamma}$, which give a possibility to investigate the operator at various values of parameters $\alpha,\beta,\gamma$.

\textbf{Remark 1.} \emph{The following constructive formulas }
\begin{equation}
\begin{array}{l}
L_{\alpha ,\beta ,\gamma} \left( {x^{1 - 2\alpha } u} \right)
\equiv x^{1 - 2\alpha } L_{1 - \alpha ,\beta ,\gamma } \left( u\right),\\
L_{\alpha ,\beta ,\gamma } \left( {y^{1 - 2\beta } u}
\right) \equiv y^{1 - 2\beta } L_{\alpha ,1 - \beta ,\gamma }\left( u \right),\\
L_{\alpha ,\beta ,\gamma } \left( {z^{1 -2\gamma} u} \right)
\equiv z^{1 - 2\gamma} L_{\alpha ,\beta ,1 - \gamma } \left( u
\right),
\end{array}
\end{equation}
\begin{equation}
\begin{array}{l}
L_{\alpha ,\beta ,\gamma } \left( {x^{1-2\alpha}y^{1 - 2\beta } u}
\right) \equiv x^{1-2\alpha}y^{1 - 2\beta } L_{1-\alpha ,1 - \beta
,\gamma } \left( u \right),\\ L_{\alpha ,\beta ,\gamma } \left(
{x^{1-2\alpha}z^{1 - 2\gamma} u} \right) \equiv x^{1-2\alpha}z^{1
- 2\gamma} L_{1-\alpha , \beta ,1-\gamma } \left( u \right), \\
L_{\alpha ,\beta ,\gamma } \left( {y^{1-2\beta}z^{1 - 2\gamma} u}
\right) \equiv y^{1-2\beta}z^{1 - 2\gamma} L_{\alpha ,1- \beta
,1-\gamma } \left( u \right),
\end{array}
\end{equation}
\begin{equation}
L_{\alpha ,\beta ,\gamma } \left( x^{1-2\alpha}y^{1-2\beta}{z^{1 -
2\gamma} u} \right) \equiv x^{1-2\alpha}y^{1-2\beta} z^{1 -
2\gamma} L_{1-\alpha ,1-\beta ,1 - \gamma } \left( u \right)
\end{equation}

\emph{are true.}

Let us prove (19). Note, equalities (17), (18) can be proved similarly.
First we find
\\$
L_{\alpha ,\beta ,\gamma } \left( x^{1-2\alpha}y^{1-2\beta}{z^{1 -
2\gamma} u} \right)
$
$$
=\left( x^{1-2\alpha}y^{1-2\beta}z^{1 -2\gamma} u\right)_{xx}+\left( x^{1-2\alpha}y^{1-2\beta}z^{1 -
2\gamma} u\right)_{yy}+\left( x^{1-2\alpha}y^{1-2\beta}z^{1 -
2\gamma} u\right)_{zz}
$$
$$
+\frac{2\alpha}{x}\left( x^{1-2\alpha}y^{1-2\beta}z^{1 -
2\gamma} u\right)_{x}+\frac{2\beta}{y}\left( x^{1-2\alpha}y^{1-2\beta}z^{1 -
2\gamma} u\right)_{y}+\frac{2\gamma}{z}\left( x^{1-2\alpha}y^{1-2\beta}z^{1 -
2\gamma} u\right)_{z}.
$$
Since
\\$
\left( x^{1-2\alpha}y^{1-2\beta}z^{1 -
2\gamma} u\right)_{xx}
$
$$
=x^{1-2\alpha}y^{1-2\beta}z^{1 -
2\gamma} u_{xx}+2(1-2\alpha)x^{-2\alpha}y^{1-2\beta}z^{1 -
2\gamma} u_x - 2\alpha(1-2\alpha)x^{-2\alpha-1}y^{1-2\beta}z^{1 -
2\gamma} u,
$$
\\$
\left( x^{1-2\alpha}y^{1-2\beta}z^{1 -
2\gamma} u\right)_{yy}
$
$$
=x^{1-2\alpha}y^{1-2\beta}z^{1 -
2\gamma} u_{yy}+2(1-2\beta)x^{1-2\alpha}y^{-2\beta}z^{1 -
2\gamma} u_y - 2\beta(1-2\beta)x^{1-2\alpha}y^{-2\beta-1}z^{1 -
2\gamma} u,
$$
\\$
\left( x^{1-2\alpha}y^{1-2\beta}z^{1 -
2\gamma} u\right)_{zz}
$
$$
=x^{1-2\alpha}y^{1-2\beta}z^{1 -
2\gamma} u_{zz}+2(1-2\gamma)x^{1-2\alpha}y^{1-2\beta}z^{ -
2\gamma} u_z - 2\gamma(1-2\gamma)x^{1-2\alpha}y^{1-2\beta}z^{ -
2\gamma-1} u,
$$
$$
\frac{2\alpha}{x}\left( x^{1-2\alpha}y^{1-2\beta}z^{1 -
2\gamma} u\right)_{x}=2\alpha x^{-2\alpha}y^{1-2\beta}z^{1 -
2\gamma} u_x + 2\alpha(1-2\alpha) x^{-2\alpha-1}y^{1-2\beta}z^{1 -
2\gamma} u,
$$
$$
\frac{2\beta}{y}\left( x^{1-2\alpha}y^{1-2\beta}z^{1 -
2\gamma} u\right)_{y}=2\beta x^{1-2\alpha}y^{-2\beta}z^{1 -
2\gamma} u_y + 2\beta(1-2\beta) x^{1-2\alpha}y^{-2\beta-1}z^{1 -
2\gamma} u,
$$
$$
\frac{2\gamma}{z}\left( x^{1-2\alpha}y^{1-2\beta}z^{1 -
2\gamma} u\right)_{z}=2\gamma x^{1-2\alpha}y^{1-2\beta}z^{ -
2\gamma} u_z + 2\gamma(1-2\gamma) x^{1-2\alpha}y^{1-2\beta}z^{ -
2\gamma-1} u,
$$
after some simplifications we get
\\$
L_{\alpha ,\beta ,\gamma } \left( x^{1-2\alpha}y^{1-2\beta}z^{1 -
2\gamma} u \right)
$
$$
=x^{1-2\alpha}y^{1-2\beta}z^{1 -
2\gamma}\left(u_{xx}+u_{yy}+u_{zz}+\frac{2(1-\alpha)}{x}u_x+\frac{2(1-\beta)}{y}u_y+\frac{2(1-\gamma)}{z}u_z\right)
$$
\\$
=x^{1-2\alpha}y^{1-2\beta}z^{1 -2\gamma}L_{1-\alpha,1-\beta,1-\gamma}(u).
$

\section{Finding fundamental solutions for the equation (1)}

Solutions of the
equation (1) we shall search as follows
\begin{equation}
u = P\left( r \right)\omega \left( {\xi ,\eta ,\zeta } \right),
\end{equation}
where
\begin{equation}
\begin{array}{l}
r^2=\left(x-x_0\right)^2+\left(y-y_0\right)^2+\left(z-z_0\right)^2,\\
r_1^2=\left(x+x_0\right)^2+\left(y-y_0\right)^2+\left(z-z_0\right)^2,\\
r_2^2=\left(x-x_0\right)^2+\left(y+y_0\right)^2+\left(z-z_0\right)^2,\\
r_3^2=\left(x-x_0\right)^2+\left(y-y_0\right)^2+\left(z+z_0\right)^2,\\
\end{array}
\end{equation}
\begin{equation}
\xi  = \displaystyle \frac{{r^2  - r_1^2 }}{{r^2 }},\eta  =
\displaystyle \frac{{r^2  - r_2^2 }}{{r^2 }},\zeta  =
\displaystyle \frac{{r^2 - r_3^2 }}{{r^2 }},\,\,\,P\left( r
\right) = \displaystyle \left( {r^2 } \right)^{ - \alpha  - \beta
- \gamma  -\displaystyle \frac{1}{2}}.
\end{equation}
Here $\forall(x,y,z)\in {\bf R}_3^+$, but
$\left(x_0,y_0,z_0\right)\in {\bf R}_3^+$ is any fixed point.
Substituting (20) into (1) we get
\begin{equation}
A_1 \omega _{\xi \xi }  + A_2 \omega _{\eta \eta } + A_3 \omega
_{\zeta \zeta }  + B_1 \omega _{\xi \eta }  + B_2 \omega _{\xi
\zeta }  + B_3 \omega _{\eta \zeta }  + C_1 \omega _\xi   + C_2
\omega _\eta   + C_3 \omega _\zeta   + D\omega  = 0,
\end{equation}
where
$$
A_1  = P\left[ {\xi _x^2  + \xi _y^2  +
\xi _z^2 } \right],\,\,A_2 = P\left[ {\eta _x^2  + \eta _y^2  +
\eta _z^2 } \right],\,\,A_3 = P\left[ {\zeta _x^2  + \zeta _y^2  +
\zeta _z^2 } \right],
$$
$$
B_1  = 2P\left[ {\xi _x \eta _x  + \xi _y \eta _y  + \xi _z \eta
_z } \right],\,\,B_2  = 2P\left[ {\xi _x \zeta _x  + \xi _y \zeta
_y + \xi _z \zeta _z } \right],\,\,B_3  = 2P\left[ {\eta _x \zeta
_x + \eta _y \zeta _y  + \eta _z \zeta _z } \right],
$$
$$
C_1  = P\left( {\xi _{xx}  + \xi _{yy}  + \xi _{zz} } \right) +
2\left( {P_x \xi _x  + P_y \xi _y  + P_z \xi _z } \right) +
P\left( {\xi _x \frac{{2\alpha }}{x} + \xi _y \frac{{2\beta }}{y}
+ \xi _z \frac{{2\gamma }}{z}} \right),
$$
$$
C_2  = P\left( {\eta _{xx}  + \eta _{yy}  + \eta _{zz} } \right) +
2\left( {P_x \eta _x + P_y \eta _y  + P_z \eta _z } \right) +
P\left( {\eta _x \frac{{2\alpha }}{x} + \eta _y \frac{{2\beta
}}{y} + \eta _z \frac{{2\gamma }}{z}} \right),
$$
$$
C_3  = P\left( {\zeta _{xx}  + \zeta _{yy}  + \zeta _{zz} }
\right) + 2\left( {P_x \zeta _x  + P_y \zeta _y  + P_z \zeta _z }
\right) + P\left( {\zeta _x \frac{{2\alpha }}{x} + \zeta _y
\frac{{2\beta }}{y} + \zeta _z \frac{{2\gamma }}{z}} \right),
$$
$$
D = P_{xx}  + P_{yy}  + P_{zz}  + P_x \frac{{2\alpha }}{x} + P_y
\frac{{2\beta }}{y} + P_z \frac{{2\gamma }}{z}.
$$
After several evaluations we find
\begin{equation}
A_1  =  - 4P\left( {r^2 } \right)^{ - 1} x^{ - 1} x_0 \xi \left(
{1 - \xi } \right),\,\,A_2 =  - 4P\left( {r^2 } \right)^{ - 1} y^{
- 1} y_0 \eta \left( {1 - \eta } \right),
\end{equation}
\begin{equation}
A_3  =  - 4P\left( {r^2 } \right)^{ - 1} z^{ - 1} z_0 \zeta \left(
{1 - \zeta } \right),\,\,B_1  = 4P\left( {r^2 } \right)^{ - 1} x^{
- 1} x_0 \xi \eta  + 4P\left( {r^2 } \right)^{ - 1} y^{ - 1} y_0
\xi \eta ,
\end{equation}
\begin{equation}
B_2  = 4P\left( {r^2 } \right)^{ - 1} x^{ - 1} x_0 \xi \zeta  +
4P\left( {r^2 } \right)^{ - 1} z^{ - 1} z_0 \xi \zeta ,\,\,B_3  =
4P\left( {r^2 } \right)^{ - 1} y^{ - 1} y_0 \eta \zeta  + 4P\left(
{r^2 } \right)^{ - 1} z^{ - 1} z_0 \eta \zeta ,
\end{equation}
\begin{equation}
\begin{array}{l}
C_1  =  - 4P\left( {r^2 } \right)^{ - 1} x^{ - 1} x_0 \left[
{2\alpha  - \left( {\alpha  + \beta  + \gamma  + \frac{1}{2} +
\alpha  + 1} \right)\xi } \right]+ 4P\left( {r^2 } \right)^{ - 1}
y^{ - 1} y_0 \beta \xi\\
+ 4P\left( {r^2 } \right)^{ - 1} z^{ - 1} z_0
\gamma \xi,\\
\end{array}
\end{equation}
\begin{equation}
\begin{array}{l} C_2  =  - 4P\left( {r^2 } \right)^{ - 1} y^{ -
1} y_0 \left[ {2\beta  - \left( {\alpha  + \beta  + \gamma  +
\frac{1}{2} + \beta  + 1} \right)\eta } \right] + 4P\left( {r^2 }
\right)^{ - 1} x^{ - 1} x_0 \alpha \eta\\ +
4P\left( {r^2 } \right)^{ - 1} z^{ - 1} z_0 \gamma \eta , \\
\end{array}
\end{equation}
\begin{equation}
\begin{array}{l} C_3  =  - 4P\left( {r^2 } \right)^{ - 1} z^{ -
1} z_0 \left[ {2\gamma  - \left( {\alpha  + \beta  + \gamma  +
\frac{1}{2} + \gamma  + 1} \right)\zeta } \right]
+ 4P\left( {r^2 } \right)^{ - 1} x^{ - 1} x_0 \alpha \zeta\\
+ 4P\left( {r^2 } \right)^{ - 1} y^{ - 1} y_0 \beta \zeta , \\
\end{array}
\end{equation}
\begin{equation}
\begin{array}{l}
D = 4P\left( {r^2 } \right)^{ - 1} x^{ - 1} x_0 \left( {\alpha  +
\beta  + \gamma  + \frac{1}{2}} \right)\alpha \, + 4P\left( {r^2 }
\right)^{ - 1} y^{ - 1} y_0 \left( {\alpha  + \beta  + \gamma  +
\frac{1}{2}} \right)\beta\\
+ 4P\left( {r^2 } \right)^{ - 1} z^{ - 1} z_0
\left( {\alpha  + \beta  + \gamma  + \frac{1}{2}} \right)\gamma . \\
\end{array}
\end{equation}
Substituting equalities (24) - (30) into (23) we obtain the
system of hypergeometric equations of Lauricella (6), which has solutions (7) - (14).

Considering (7) - (14), from (20) we obtain
eight FSs of the equation (1)
\begin{equation}
q_1 \left( {x,y,z;x_0 ,y_0 ,z_0 } \right) = k_1 \left( {r^2 }
\right)^{ - \alpha  - \beta  - \gamma  - \frac{1}{2}} F_A^{\left(
3 \right)} \left( {\alpha  + \beta  + \gamma  + \frac{1}{2};\alpha
,\beta ,\gamma ;2\alpha ,2\beta ,2\gamma ;\xi ,\eta ,\zeta }
\right),
\end{equation}
\begin{equation}
\begin{array}{l}
q_2 \left( {x,y,z;x_0 ,y_0 ,z_0 } \right) \\ = k_2 \left( {r^2 }
\right)^{\alpha  - \beta  - \gamma  - \displaystyle
\frac{3}{2}}x^{1 - 2\alpha } x_0^{1 - 2\alpha } F_A^{\left( 3
\right)} \left( { - \alpha  + \beta  + \gamma  + \displaystyle
\frac{3}{2};1 - \alpha ,\beta ,\gamma ;2 - 2\alpha , 2\beta
,2\gamma ;\xi ,\eta ,\zeta } \right),
\end{array}
\end{equation}
\begin{equation}
\begin{array}{l}
q_3 \left( {x,y,z;x_0 ,y_0 ,z_0 } \right) \\ = k_3 \left( {r^2 }
\right)^{ - \alpha  + \beta  - \gamma  - \displaystyle
\frac{3}{2}}y^{1 - 2\beta } y_0^{1 - 2\beta } F_A^{\left( 3
\right)} \left( {\alpha  - \beta + \gamma  + \displaystyle
\frac{3}{2};\alpha ,1 - \beta , \gamma ;2\alpha ,2 - 2\beta
,2\gamma ;\xi ,\eta ,\zeta } \right),
\end{array}
\end{equation}
\begin{equation}
\begin{array}{l}
q_4 \left( {x,y,z;x_0 ,y_0 ,z_0 } \right) \\ = k_4 \left( {r^2 }
\displaystyle \right)^{ - \alpha  - \beta  + \gamma  -
\displaystyle \frac{3}{2}}z^{1 - 2\gamma } z_0^{1 - 2\gamma }
F_A^{\left( 3 \right)} \left( {\alpha + \beta  - \gamma  +
\displaystyle \frac{3}{2};\alpha ,\beta ,1 - \gamma ;2\alpha
,2\beta ,2 - 2\gamma ;\xi ,\eta ,\zeta } \right),
\end{array}
\end{equation}
\begin{equation}
\begin{array}{l}
q_5 \left( {x,y,z;x_0 ,y_0 ,z_0 } \right) = k_5 \left( {r^2 }
\right)^{\alpha  + \beta  -\gamma  - \displaystyle \frac{5}{2}}  \\
\times x^{1 - 2\alpha } y^{1 - 2\beta } x_0^{1 - 2\alpha } y_0^{1 -
2\beta } F_A^{\left( 3 \right)} \left( { - \alpha  - \beta  +
\gamma  + \displaystyle \frac{5}{2};1 -\alpha ,1 - \beta ,
\gamma ;2 - 2\alpha ,2 - 2\beta ,2\gamma ;\xi ,\eta ,\zeta } \right), \\
\end{array}
\end{equation}
\begin{equation}
\begin{array}{l}
q_6 \left( {x,y,z;x_0 ,y_0 ,z_0 } \right) = k_6 \left( {r^2 }
\right)^{\alpha  -\beta  + \gamma  - \displaystyle \frac{5}{2}}
\\ \times x^{1 - 2\alpha } z^{1 - 2\gamma } x_0^{1 - 2\alpha }
z_0^{1 - 2\gamma } F_A^{\left( 3 \right)} \left( { - \alpha  +
\beta  -\gamma  + \displaystyle \frac{5}{2};1 -\alpha ,\beta ,1 -
\gamma ;2 - 2\alpha ,2\beta ,2 - 2\gamma ;\xi ,\eta ,\zeta } \right), \\
\end{array}
\end{equation}
\begin{equation}
\begin{array}{l}
q_7 \left( {x,y,z;x_0 ,y_0 ,z_0 } \right) = k_7 \left( {r^2 }
\right)^{ - \alpha  +\beta  + \gamma  - \displaystyle \frac{5}{2}}  \\
\times y^{1 - 2\beta } z^{1 - 2\gamma } y_0^{1 - 2\beta } z_0^{1 -
2\gamma } F_A^{\left( 3 \right)} \left( {\alpha  - \beta  - \gamma
+ \displaystyle \frac{5}{2};\alpha ,1 -  \beta ,1 - \gamma ;2\alpha ,2 -
2\beta ,2 - 2\gamma ;\xi ,\eta ,\zeta } \right), \\
\end{array}
\end{equation}
\begin{equation}
\begin{array}{l}
q_8 \left( {x,y,z;x_0 ,y_0 ,z_0 } \right) = k_8 \left( {r^2 }
\right)^{\alpha  + \beta  + \gamma  - \displaystyle \frac{7}{2}}
x^{1 - 2\alpha } y^{1 - 2\beta } z^{1 - 2\gamma }x_0^{1 - 2\alpha } y_0^{1 - 2\beta } z_0^{1 - 2\gamma }  \\
\times F_A^{\left( 3 \right)} \left( { - \alpha  - \beta  - \gamma
+ \displaystyle \frac{7}{2};1 - \alpha ,1 - \beta ,1 - \gamma ;2 -
2\alpha ,2 - 2\beta ,2 - 2\gamma ;\xi ,\eta ,\zeta }
\right). \\
\end{array}
\end{equation}
Here $k_i\,\,(i=\overline{1,8})$ are constants. They will be determined when we solve boundary-value problems.

Let us show that the found solutions (31) - (38) have a singularity.

We choose a solution $ q_1
\left( {x,y,z;x_0 ,y_0 ,z_0 } \right)$. For this aim we use
expansion for hypergeometric function of Lauricella (15).  As a result solution (31) can be written as follows
\begin{equation}
\begin{array}{l}
q_1 \left( {x,y,z;x_0 ,y_0 ,z_0 } \right) = k_1 \left( {r^2 }
\right)^{ - \alpha  - \beta  - \gamma  - \displaystyle
\frac{1}{2}} \displaystyle{\sum\limits_{l,m,n = 0}^\infty  {}} \displaystyle
\frac{{\left( {\alpha + \beta  + \gamma  + \displaystyle
\frac{1}{2}} \right)_{l + m + n} \left( \alpha \right)_{l + m}
\left( \beta  \right)_{l + n} \left( \gamma \right)_{m + n}
}}{{\left( {2\alpha } \right)_{l + m} \left( {2\beta } \right)_{l
+ n} \left( {2\gamma } \right)_{m + n} l!m!n!}} \\
\times \left( {1 - \displaystyle \frac{{r_1^2 }}{{r^2 }}}
\right)^{l + m} \left( {1 - \displaystyle \frac{{r_2^2 }}{{r^2 }}}
\right)^{l + n} \left( {1 - \displaystyle \frac{{r_3^2 }}{{r^2 }}} \right)^{m + n}  \\
\times {}_2F_1 \left( {\alpha  + \beta  + \gamma  + \displaystyle
\frac{1}{2} + l + m,\alpha  + l + m;2\alpha  + l + m;1 - \displaystyle \frac{{r_1^2 }}{{r^2 }}} \right) \\
\times {}_2F_1 \left( {\alpha  + \beta  + \gamma  + \displaystyle
\frac{1}{2} + l + m + n,\beta  +l + n;2\beta  + l + n;1 - \displaystyle \frac{{r_2^2 }}{{r^2 }}} \right) \\
\times {}_2F_1 \left( {\alpha  + \beta  + \gamma  + \displaystyle
\frac{1}{2} + l + m + n,\gamma  +m + n;2\gamma  + m + n;1 -
\displaystyle \frac{{r_3^2 }}{{r^2 }}} \right).
\end{array}
\end{equation}
Using formula (5), we rewrite (39) as
\begin{equation}
q_1 \left( {x,y,z;x_0 ,y_0 ,z_0 } \right) = k_1 r^{ - 1} \left(
{r_1^2 } \right)^{ - \alpha } \left( {r_2^2 } \right)^{ - \beta }
\left( {r_3^2 } \right)^{ - \gamma } f\left( {r^2 ,r_1^2 ,r_2^2
,r_3^2 } \right),
\end{equation}
where
\begin{equation}
\begin{array}{l}
f\left( {r^2 ,r_1^2 ,r_2^2 ,r_3^2 } \right) =
\displaystyle{\sum\limits_{l,m,n = 0}^\infty  {} \displaystyle
\frac{{\left( {\alpha  + \beta  + \gamma  + \displaystyle
\frac{1}{2}} \right)_{l + m + n} \left( \alpha \right)_{l + m}
\left( \beta \right)_{l + n} \left( \gamma \right)_{m + n}
}}{{\left( {2\alpha } \right)_{l + m} \left( {2\beta } \right)_{l
+ n} \left( {2\gamma } \right)_{m + n} l!m!n!}}} \\
\times \left(\displaystyle {\frac{{r^2 }}{{r_1^2 }} - 1} \right)^{l
+ m} \left( \displaystyle {\frac{{r^2 }}{{r_2^2 }} - 1} \right)^{l
+ n} \left(\displaystyle {\frac{{r^2 }}{{r_3^2 }} - 1} \right)^{m + n}  \\
\times {}_2F_1 \left( {\alpha  - \beta  - \gamma  - \displaystyle \frac{1}{2},
\alpha  + l + m;2\alpha  +l + m;1 - \displaystyle \frac{{r^2 }}{{r_1^2 }}} \right) \\
\times {}_2F_1 \left( {\beta  - \alpha  - \gamma  - \displaystyle
\frac{1}{2} - m,\beta  + l + n;2\beta  + l + n;1 - \displaystyle \frac{{r^2 }}{{r_2^2 }}} \right) \\
\times {}_2F_1 \left( {\gamma  - \alpha  - \beta  - \displaystyle
\frac{1}{2} - l,\gamma  + m + n;2\gamma  + m + n;1 - \displaystyle
\frac{{r^2 }}{{r_3^2 }}} \right).
\end{array}
\end{equation}

Below we show that $f\left( {r^2 ,r_1^2 ,r_2^2 ,r_3^2 } \right)$ will be constant at $r\rightarrow 0$.

For this aim we use an equality (4). Then we get
$$
\begin{array}{l}
\left. {{}_2F_1 \left( {\alpha  - \beta  - \gamma  - \displaystyle
\frac{1}{2},\alpha  + l + m;2\alpha  + l + m;1 - \displaystyle
\frac{{r^2 }}{{r_1^2 }}} \right)} \right|_{r  =0}\\
=\displaystyle {\frac{{\Gamma \left( {\beta  + \gamma +
\displaystyle \frac{1}{2}} \right)\Gamma \left( {2\alpha+l+m }
\right)}}{{\Gamma \left( \alpha
\right)\Gamma \left( {\alpha  + \beta  + \gamma  + \displaystyle
\frac{1}{2}+l+m} \right)}}}.
\end{array}
$$
Considering first formula in (3), we obtain
\begin{equation}
\begin{array}{l}
\left. {{}_2F_1 \left( {\alpha  - \beta  - \gamma  - \displaystyle
\frac{1}{2},\alpha  + l + m;2\alpha  + l + m;1 - \displaystyle
\frac{{r^2 }}{{r_1^2 }}} \right)} \right|_{r  =0}\\
=\displaystyle {\frac{{\Gamma \left( {\beta  + \gamma +
\displaystyle \frac{1}{2}} \right)\Gamma \left( {2\alpha }
\right)\left( {2\alpha } \right)_{l + m} }}{{\Gamma \left( \alpha
\right)\Gamma \left( {\alpha  + \beta  + \gamma  + \displaystyle
\frac{1}{2}} \right)\left( {\alpha  + \beta  + \gamma +
\displaystyle \frac{1}{2}} \right)_{l + m} }}}.
\end{array}
\end{equation}
Similarly we get
\begin{equation}
\begin{array}{l}
\left. {{}_2F_1 \left( {\beta  - \alpha  - \gamma  - \displaystyle
\frac{1}{2} - m,\beta  + l + n;2 \beta  + l + n;1 - \displaystyle
\frac{{r^2 }}{{r_2^2 }}} \right)} \right|_{r  = 0}  \\
= \displaystyle {\frac{{\Gamma \left( {2\beta } \right)\Gamma
\left( {\alpha  + \gamma  + \displaystyle \frac{1}{2}}
\right)}}{{\Gamma \left( \beta  \right)\Gamma \left( {\alpha  +
\beta  + \gamma  + \displaystyle \frac{1}{2}} \right)}}
\displaystyle \frac{{\left( {2\beta } \right)_{l + n} \left( {
\alpha + \gamma  + \displaystyle \frac{1}{2}} \right)_m }}{{\left(
{\alpha  + \beta  + \gamma  + \displaystyle \frac{1}{2}}
\right)_{l + m + n} }}},
\end{array}
\end{equation}
\begin{equation}
\begin{array}{l}
\left. {{}_2F_1 \left( {\gamma  - \alpha  - \beta  - \displaystyle
\frac{1}{2} - l,\gamma  + m + n; 2\gamma  + m + n;1 -
\displaystyle \frac{{r^2 }}{{r_3^2 }}} \right)} \right|_{r  = 0}  \\
= \displaystyle{\frac{{\Gamma \left( {2\gamma } \right)\Gamma
\left( {\alpha  + \beta  + \displaystyle \frac{1}{2}}
\right)}}{{\Gamma \left( \gamma  \right)\Gamma \left( {\alpha  +
\gamma  + \beta  + \displaystyle \frac{1}{2}} \right)}}
\displaystyle \frac{{\left( {2\gamma } \right)_{m + n} \left(
{\alpha + \beta  + \displaystyle \frac{1}{2}} \right)_l }}{{\left(
{\alpha  + \gamma  + \beta  + \displaystyle \frac{1}{2}}
\right)_{l + m + n} }}}.
\end{array}
\end{equation}

Taking (42) - (44) into account from (41) at $r\rightarrow 0$ we have
\begin{equation}
\begin{array}{l}
f\left( {0,r_1^2 ,r_2^2 ,r_3^2 } \right) =
\displaystyle{\frac{{\Gamma \left( {2\alpha } \right)\Gamma \left(
{2\beta } \right)\Gamma \left( {2\gamma } \right)\Gamma \left( {
\alpha  + \beta  + \displaystyle \frac{1}{2}} \right)\Gamma \left(
{\alpha  + \gamma  + \displaystyle \frac{1}{2}} \right)\Gamma
\left( {\beta  + \gamma  + \displaystyle \frac{1}{2}}
\right)}}{{\Gamma \left( \alpha \right)\Gamma \left( \beta
\right)\Gamma \left( \gamma \right)\Gamma ^3 \left(
{\alpha  + \beta  + \gamma  + \displaystyle \frac{1}{2}} \right)}}} \\
\cdot \displaystyle{\sum\limits_{l,m,n = 0}^\infty {}
\displaystyle \frac{{\left( {\alpha  + \gamma  + \displaystyle
\frac{1}{2}} \right)_m \left( {\alpha  + \beta  + \displaystyle
\frac{1}{2}} \right)_l \left( \alpha \right)_{l + m} \left( \beta
\right)_{l + n} \left( \gamma \right)_{m + n} }}{{ \left( {\alpha
+ \gamma  + \beta  + \displaystyle \frac{1}{2}} \right)_{l + m +
n} \left( {\alpha  + \beta  + \gamma  + \displaystyle \frac{1}{2}}
\right)_{l + m} l!m!n!}}}.
\end{array}
\end{equation}

We use formulas in (3) to simplify (45). Since
$$
\begin{array}{l}
\displaystyle{\sum\limits_{l,m,n = 0}^\infty {}
\displaystyle \frac{{\left( {\alpha  + \gamma  + \displaystyle
\frac{1}{2}} \right)_m \left( {\alpha  + \beta  + \displaystyle
\frac{1}{2}} \right)_l \left( \alpha \right)_{l + m} \left( \beta
\right)_{l + n} \left( \gamma \right)_{m + n} }}{{ \left( {\alpha
+ \gamma  + \beta  + \displaystyle \frac{1}{2}} \right)_{l + m +
n} \left( {\alpha  + \beta  + \gamma  + \displaystyle \frac{1}{2}}
\right)_{l + m} l!m!n!}}}\\
=\displaystyle{\sum\limits_{l,m,n = 0}^\infty {}
\displaystyle \frac{{\left( {\alpha  + \gamma  + \displaystyle
\frac{1}{2}} \right)_m \left( {\alpha  + \beta  + \displaystyle
\frac{1}{2}} \right)_l \left( \alpha \right)_l\left( \alpha+l \right)_m \left( \beta
\right)_l \left( \beta+l\right)_n\left( \gamma \right)_m\left( \gamma+m \right)_n }}{{ \left( {\alpha
+ \gamma  + \beta  + \displaystyle \frac{1}{2}+l+m} \right)_n\left(\left( {\alpha  + \beta  + \gamma  + \displaystyle \frac{1}{2}}
\right)_{l + m}\right)^2 l!m!n!}}}\\
=\displaystyle{\sum\limits_{l,m = 0}^\infty {}
\displaystyle \frac{{\left( {\alpha  + \gamma  + \displaystyle
\frac{1}{2}} \right)_m \left( {\alpha  + \beta  + \displaystyle
\frac{1}{2}} \right)_l \left( \alpha \right)_l\left( \alpha+l \right)_m \left( \beta
\right)_l \left( \gamma \right)_m}}{{ \left(\left( {\alpha  + \beta  + \gamma  + \displaystyle \frac{1}{2}}
\right)_{l + m}\right)^2 l!m!}}}\cdot
\displaystyle{\sum\limits_{l,m = 0}^\infty {}
\displaystyle{\frac{{\left( \beta+l\right)_n\left( \gamma+m \right)_n }}{{ \left( {\alpha
+ \gamma  + \beta  + \displaystyle \frac{1}{2}+l+m} \right)_n n!}}}}\\
=\displaystyle{\frac{\Gamma\left(\alpha+\beta+\gamma+\displaystyle{\frac{1}{2}}\right)\Gamma\left(\alpha+\displaystyle{\frac{1}{2}}\right)}
{\Gamma\left(\alpha+\gamma+\displaystyle{\frac{1}{2}}\right)\Gamma\left(\alpha+\beta+\displaystyle{\frac{1}{2}}\right)}
\displaystyle{\sum\limits_{l = 0}^\infty {}
\displaystyle{\frac{(\alpha)_l(\beta)_l}{\left(\alpha+\beta+\gamma+\displaystyle{\frac{1}{2}}\right)_l l!}
\sum\limits_{m = 0}^\infty {}\displaystyle{\frac{(\gamma)_m(\alpha+l)_m}{\left(\alpha+\beta+\gamma+\displaystyle{\frac{1}{2}}+l\right)_m m!}
}}}}\\
=\displaystyle{\frac{\Gamma^2\left(\alpha+\beta+\gamma+\displaystyle{\frac{1}{2}}\right)\Gamma\left(\alpha+\displaystyle{\frac{1}{2}}\right)
\Gamma\left(\beta+\displaystyle{\frac{1}{2}}\right)}
{\Gamma^2\left(\alpha+\beta+\displaystyle{\frac{1}{2}}\right)\Gamma\left(\alpha+\gamma+\displaystyle{\frac{1}{2}}\right)
\Gamma\left(\beta+\gamma+\displaystyle{\frac{1}{2}}\right)}
\displaystyle{\sum\limits_{l = 0}^\infty {}
\displaystyle{\frac{(\alpha)_l(\beta)_l}{\left(\alpha+\beta+\gamma+\displaystyle{\frac{1}{2}}\right)_l l!}
}}}\\
= \displaystyle{\frac{{\Gamma ^2 \left( {\alpha  + \gamma  + \beta
+ \displaystyle \frac{1}{2}} \right)\sqrt \pi  }}{{\Gamma \left(
{\alpha  + \beta  + \displaystyle \frac{1}{2}} \right)\Gamma
\left( {\alpha  + \gamma  + \displaystyle \frac{1}{2}}
\right)\Gamma \left( {\beta  + \gamma  + \displaystyle
\frac{1}{2}} \right)}}},
\end{array}
$$
from (45) follows
\begin{equation}
f\left( {0,r_1^2 ,r_2^2 ,r_3^2 } \right) = \displaystyle
\frac{{\Gamma \left( {2\alpha } \right)\Gamma \left( {2\beta }
\right)\Gamma \left( {2\gamma } \right)\sqrt \pi }}{{\Gamma \left(
\alpha \right)\Gamma \left( \beta  \right)\Gamma \left( \gamma
\right)\Gamma \left( {\alpha  + \beta  + \gamma  + \displaystyle
\frac{1}{2}} \right)}}.
\end{equation}
Expressions (46) and (40) gives us possibility to conclude that the solution $q_1 \left( {x,y,z;x_0
,y_0 ,z_0 } \right)$ reduce to infinity of the order $r^{ - 1}$ at
$r  \to 0$. Similarly it is possible to be convinced that
solutions $q_i \left( {x,y,z;x_0 ,y_0 ,z_0 } \right),\,\,i =
2,3,...,8$ also reduce to infinity of the order $r^{ - 1}$ when $r
\to 0$.

It is directly checkable that constructed functions (31) -
(38) possess properties
$$
x^{2\alpha}\frac{\partial}{\partial x}q_1 \bigr |_{x=0}=0,\,\,\,
y^{2\beta}\frac{\partial}{\partial y}q_1 \bigr |_{y=0}=0,\,\,\,
z^{2\gamma}\frac{\partial}{\partial z}q_1 \bigr |_{z=0}=0,
$$
$$
q_2 \bigr |_{x=0}=0,\,\,\, y^{2\beta}\frac{\partial}{\partial
y}q_2 \bigr |_{y=0}=0,\,\,\, z^{2\gamma}\frac{\partial}{\partial
z}q_2 \bigr |_{z=0}=0,
$$
$$
x^{2\alpha}\frac{\partial}{\partial x}q_3 \bigr |_{x=0}=0,\,\,\,
q_3 \bigr |_{y=0}=0,\,\,\, z^{2\gamma}\frac{\partial}{\partial
z}q_3 \bigr |_{z=0}=0,
$$
$$
x^{2\alpha}\frac{\partial}{\partial x}q_4 \bigr |_{x=0}=0,\,\,\,
y^{2\beta}\frac{\partial}{\partial y}q_4 \bigr |_{y=0}=0,\,\,\,
q_4 \bigr |_{z=0}=0,
$$
$$
q_5 \bigr |_{x=0}=0,\,\,\, q_5 \bigr |_{y=0}=0,\,\,\,
z^{2\gamma}\frac{\partial}{\partial z}q_5 \bigr |_{z=0}=0,
$$
$$
q_6 \bigr |_{x=0}=0,\,\,\, y^{2\beta}\frac{\partial}{\partial
y}q_6 \bigr |_{y=0}=0,\,\,\, q_6 \bigr |_{z=0}=0,
$$
$$
x^{2\alpha}\frac{\partial}{\partial x}q_7 \bigr |_{x=0}=0,\,\,\,
q_7 \bigr |_{y=0}=0,\,\,\, q_7 \bigr |_{z=0}=0,
$$
$$
q_8 \bigr |_{x=0}=0, q_8 \bigr |_{y=0}=0, q_8  \bigr |_{z=0}=0.
$$

Let be convinced that
\begin{equation}
x^{2\alpha}\frac{\partial}{\partial x}q_1 \bigr |_{x=0}=0.
\end{equation}
Using formula of differentiation for Lauricella HF of three variables (16) we have
$$
\begin{array}{l}
 \displaystyle{\frac{\partial }{{\partial x}}q_1 \left( {x,y,z;x_0 ,y_0 ,z_0 } \right) =  - 2\left( {x - x_0 } \right)\left( {\alpha  + \beta  + \gamma  + \frac{1}{2}} \right)k_1 \left( {r^2 } \right)^{ - \alpha  - \beta  - \gamma  - \frac{3}{2}}  }  \\
\displaystyle{  \times \left[ {F_A^{(3)} \left( {\alpha  + \beta  + \gamma  + \frac{3}{2};\alpha ,\beta ,\gamma ;2\alpha ,2\beta ,2\gamma ;\xi ,\eta ,\zeta } \right)  } \right.} \\
\displaystyle{  + \frac{\xi }{{2}} F_A^{(3)} \left( {\alpha  + \beta  + \gamma  + \frac{3}{2};\alpha  + 1,\beta ,\gamma ;2\alpha  + 1,2\beta ,2\gamma ;\xi ,\eta ,\zeta } \right)  } \\
\displaystyle{+ \frac{\eta }{{2}}F_A^{(3)} \left( {\alpha  + \beta  + \gamma  + \frac{3}{2};\alpha ,\beta  + 1,\gamma ;2\alpha ,2\beta  + 1,2\gamma ;\xi ,\eta ,\zeta } \right) }  \\
 \displaystyle{\left. {  \frac{\zeta }{{2}}F_A^{(3)} \left( {\alpha  + \beta  + \gamma  + \frac{3}{2};\alpha ,\beta ,\gamma  + 1;2\alpha ,2\beta ,2\gamma  + 1;\xi ,\eta ,\zeta } \right)} \right] }  \\
\displaystyle{  - 4x_0 \left( {\alpha  + \beta  + \gamma  + \frac{1}{2}} \right)k_1 \left( {r^2 } \right)^{ - \alpha  - \beta  - \gamma  - \frac{3}{2}} F_A^{(3)} \left( {\alpha  + \beta  + \gamma  + \frac{3}{2};\alpha ,\beta ,\gamma ;2\alpha ,2\beta ,2\gamma ;\xi ,\eta ,\zeta } \right).} \\
 \end{array}
$$
From the last equality one can easily be convinced that (47) is true.

\textbf{Remark.} \emph{Afore-mentioned properties of found fundamental solutions will be used at solving various boundary-value problems for the equation (1). For instance, fundamental solution (31) with its properties will be used at solving problem N for the equation (1).}

\subsection*{Acknowledgement.}
The authors are grateful to Professor R.Gilbert for his kind interest and given materials on a theme of the paper.

\end{document}